\documentclass[sigconf]{acmart}

\usepackage{booktabs}
\usepackage{caption}
\usepackage{amsmath}

\newcommand{\Pop}{\mathcal{P}} 
\usepackage{enumitem}
\usepackage[linesnumbered, ruled, vlined]{algorithm2e}
\usepackage{xcolor}
\usepackage{stfloats} 
\SetAlCapNameFnt{\bfseries}
\SetAlCapFnt{\bfseries}
\SetNlSty{textbf}{\color{black}}{}
\SetAlgoNlRelativeSize{-1}
\SetKwInput{KwInput}{Input}
\SetKwInput{KwOutput}{Output}
\SetKw{KwTo}{to}
\SetKw{KwBy}{by}
\SetKw{KwRet}{return}

\usepackage{multirow}
\usepackage{booktabs} 
\usepackage{caption}  
\usepackage{graphicx} 

\usepackage{xcolor}     
\usepackage{tikz}       
\usepackage{amssymb}    
\usepackage{makecell}

\newcommand{\circlesize}{0.6ex}
\newcommand{\coloredcircle}[1]{%
  \tikz[baseline=-0.6ex]\fill[#1] (0,0) circle (\circlesize);%
}
\newcommand{\redcircle}{\coloredcircle{red}}
\newcommand{\yellowcircle}{\coloredcircle{orange!85!yellow}}
\newcommand{\greencircle}{\coloredcircle{green!65!black}}
\newcommand{\bluecircle}{\coloredcircle{blue!70!black}}

\usepackage{pifont}
\newcommand{\XSolidBrush}{\ding{55}}  
\newcommand{\Checkmark}{\ding{51}}   
\usepackage{subcaption}

\acmYear{2026}
\acmConference[DAC '26]{63rd ACM/IEEE Design Automation Conference}{July 26--29, 2026}{Long Beach, CA, USA}
\acmBooktitle{Proceedings of the 63rd ACM/IEEE Design Automation Conference (DAC '26), July 26--29, 2026, Long Beach, CA, USA}
\acmISBN{978-1-4503-9857-0}
\acmDOI{10.1145/3770743.3804350}  

\renewcommand{\acmISBN}{978-1-4503-9857-0}
\renewcommand{\acmDOI}{10.1145/3770743.3804350}  

\setcopyright{none}  

\begin{document}

\title{PRISM: Dynamic Primitive-Based Forecasting for Large-Scale GPU Cluster Workloads}

\author{Xin Wu$^1$, Fei Teng$^{1,2}$$^*$, Xingwang Li$^1$, Bin Zheng$^3$, Qiang Duan$^4$}
\affiliation{%
    \institution{$^1$Southwest Jiaotong University \hspace{0.3em}$^2$Engineering Research Center of Sustainable Urban Intelligent Transportation}
    \institution{$^3$Beijing University of Posts and Telecommunications  \hspace{0.3em}$^4$The Pennsylvania State University}
   \country{}
}
\affiliation{%
    \institution{$^*$Corresponding author: fteng@swjtu.edu.cn}
 \country{}
}

\renewcommand{\shortauthors}{Xin Wu et al.}


\begin{abstract}
Accurately forecasting GPU workloads is essential for AI infrastructure, enabling efficient scheduling, resource allocation, and power management. Modern workloads are highly volatile, multiple periodicity, and heterogeneous, making them challenging for traditional predictors. We propose PRISM, a primitive-based compositional forecasting framework combining dictionary-driven temporal decomposition with adaptive spectral refinement. This dual representation extracts stable, interpretable workload signatures across diverse GPU jobs. Evaluated on large-scale production traces, PRISM achieves state-of-the-art results. It significantly reduces burst-phase errors, providing a robust, architecture-aware foundation for dynamic resource management in GPU-powered AI platforms.
\end{abstract}



\maketitle

\section{INTRODUCTION}
\begin{figure}
\centering
\includegraphics[width=0.92\columnwidth]{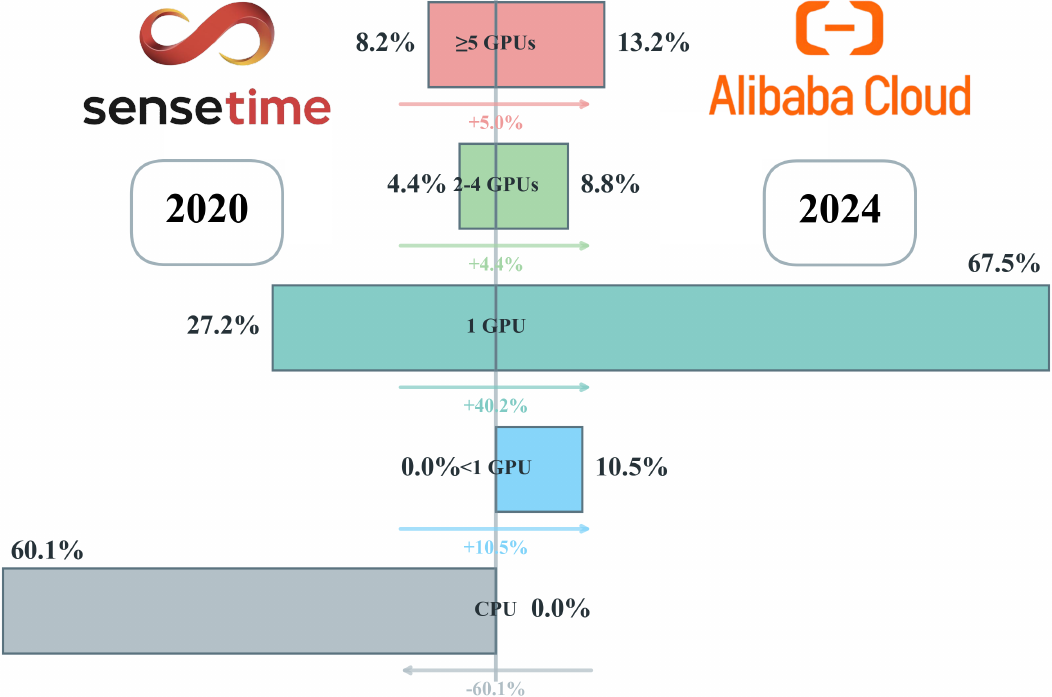} 
\caption{Contrast in resource request profiles between two large-scale production clusters. Workload profiles show significant differences. The 2020 cluster profile included a 60.1\% CPU-based majority \cite{Hu2021SC}. In contrast, the 2024 cluster profile is GPU-centric and polarized, centering on single-GPU requests (67.5\%) while expanding toward coarse-grained (13.2\%) and fine-grained (10.5\%) allocations \cite{duan2025gfs}.}
\label{fig_intro_1}
\end{figure}

Effective resource management is a cornerstone of modern, large-scale GPU AI infrastructure \cite{2020Kinoshita, Alcaide2019DATE, 2025HuangLLSM, 2025XuASPDAC}. The rapid growth of deep learning services has introduced unprecedented workload dynamics, ranging from fine-grained fractional-GPU jobs to coarse-grained multi-GPU allocations \cite{2020Kim, 2023XieICCAD}. Such heterogeneity creates fundamental conflicts with static resource allocation policies, intensifying the trade-off between resource efficiency and quality-of-service guarantees. Fig. \ref{fig_intro_1} illustrates this polarization in GPU demand in a production cluster. These challenges highlight the urgent need for adaptive, architecture-aware methodologies that enable intelligent scheduling \cite{2014DATELee, 2024Taufique, 2025LeeASPDAC, 2025Falahati}, and system-level optimization for next-generation AI platforms \cite{2025Zhao, 2025Zhong}.

To achieve such adaptive scheduling, accurate workload forecasting is indispensable. Yet, classical statistical models such as ARIMA \cite{6881647} assume stationarity, making them poorly suited to the highly stochastic behavior of GPU workloads \cite{2023Bi,2025Mo}. As a result, they struggle with extreme volatility and nonlinear periodicities, often incurring large errors. AI-based approaches have therefore emerged, including Transformer-style architectures such as Autoformer \cite{wu2021autoformer}, TimesNet \cite{wu2023timesnet}, and MetaEformer \cite{2025kddMetaEformer}, as well as simplified MLP and linear models such as TimeXer \cite{wang2024timexer} and Dlinear \cite{zeng2023transformers}. However, these methods share a core limitation: they function as monolithic global learners. They treat aggregated demand, which is actually a dynamic superposition of heterogeneous micro-entities \cite{10912752}, as if it were a homogeneous signal. By attempting to model a nonexistent "average user", they regress to the mean and systematically smooth out critical peaks. This inability to disentangle heterogeneity ultimately undermines their reliability for risk-aware resource scheduling \cite{2017Shieh, 2024Morchdi, 2025Zhao}.

\begin{figure}
\centering
\includegraphics[width=0.98\columnwidth]{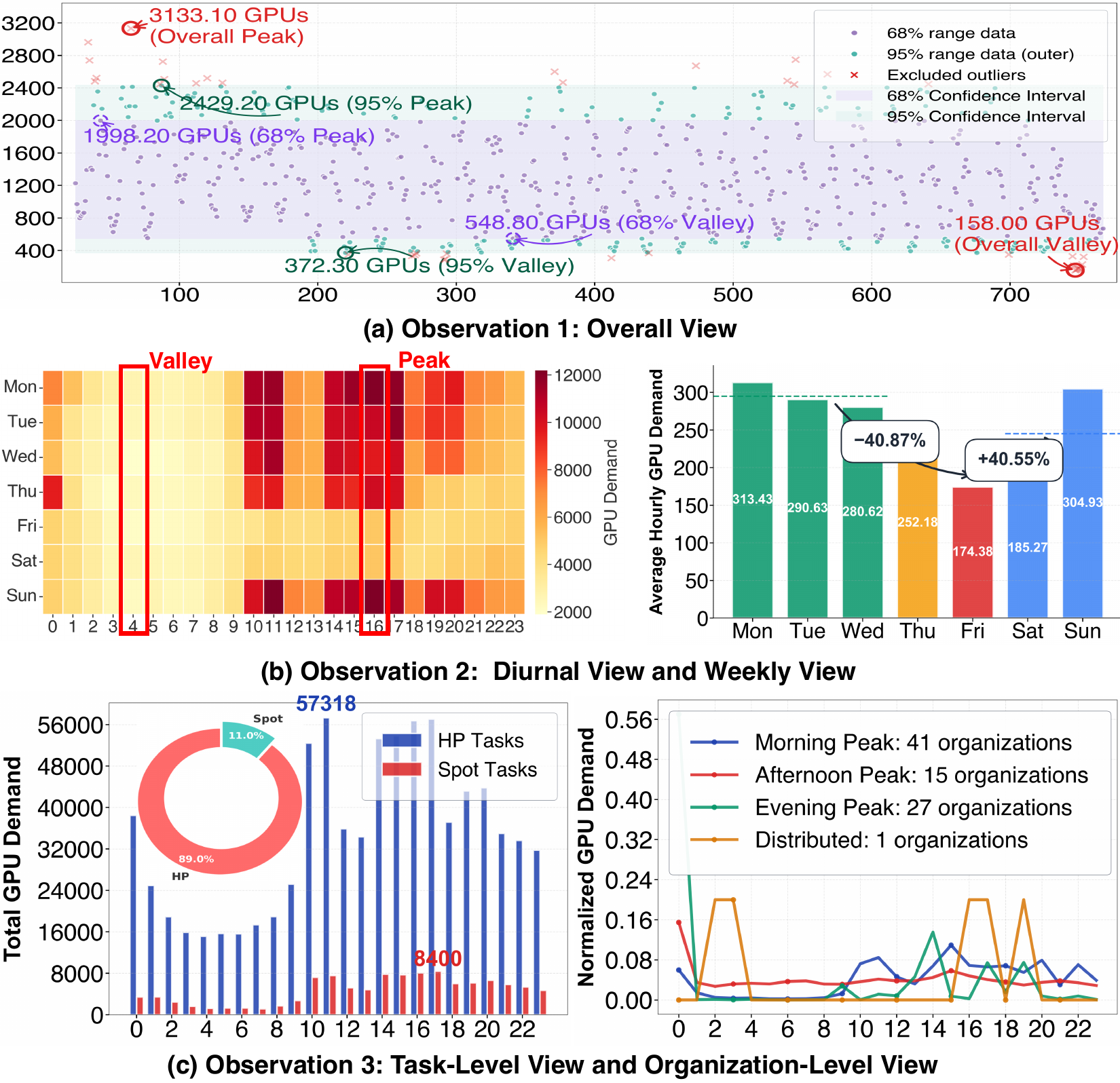} 
\caption{A multi-faceted analysis revealing GPU demand volatility, periodicity, and heterogeneity.}
\label{fig_intro_2}
\end{figure}

In this work, we present \textbf{PRISM} (\underline{\textbf{Pr}}imitive-based \underline{\textbf{I}}nterpretable \underline{\textbf{S}}ignal \underline{\textbf{M}}ixture), a novel solution that provides a compositional and interpretable framework for modeling large-scale GPU workloads. (1) \textbf{Decomposability:} The framework decomposes the complex aggregated signal into a set of interpretable base units that reflect foundational behaviors. (2) \textbf{Heterogeneity Modeling:} Instead of learning a single average pattern, the model reconstructs the high-level heterogeneity of the multi-tenant environment \cite{2025Yu} through the dynamic combination of these primitives, adapting to the unique patterns of diverse user groups. As illustrated in Fig. \ref{fig_intro_2}, our methodology is predicated on three key observations:
\begin{itemize}
 \item \textbf{Observation 1:} The inherent conflict between extreme workload volatility and static resource allocation. Observational data reveal that the non-stationary nature of the total cluster GPU demand results in an exceptionally wide dynamic range, with a peak-to-trough ratio of up to \underline{\textbf{19.83x}}. The regular dynamic range (95\% confidence interval) is also \underline{\textbf{6.52x}}. This severe volatility traps static allocation strategies in an inescapable provisioning dilemma: over-provisioning based on peaks leads to immense cost waste, while under-provisioning based on the mean induces service-level agreement (SLA) violations during bursts.

 \item \textbf{Observation 2:}  The workload is not merely stationary; it results from a non-linear superposition of periodic patterns occurring at multiple time scales, including diurnal and weekly cycles. Notably, there is a distinctive weekly pattern characterized by a "Friday dip and weekend rebound." This complexity presents a significant challenge for predictive models that assume a single, fixed periodicity.

 
\item \textbf{Observation 3:} Significant heterogeneity exists in user behavior patterns. The aggregated workload emerges from the dynamic superposition of numerous micro-entities exhibiting diverse behaviors. This heterogeneity is evident along two dimensions: task type, distinguishing high-priority (HP) from low-priority Spot tasks, and user temporal patterns, differentiating users with distinct active periods, such as morning and afternoon peaks. Such pronounced heterogeneity makes any monolithic global model attempting to represent an "average user" fundamentally ill-posed.
\end{itemize}

In response to these challenges, PRISM abandons the traditional single global model approach and adopts a compositional forecasting paradigm through a dual-component encoder. One component, the primitive dictionary decomposition module, learns basic temporal patterns, such as diurnal cycles. The other component, the adaptive spectral refinement module, improves the signal representation in the frequency domain. A dynamic mixture network then combines these decomposed and refined features to reconstruct complex and heterogeneous workload patterns adaptively.


The contributions of this paper are summarized as follows:
\begin{itemize}
    \item We propose PRISM, a compositional forecasting framework designed to handle the extreme volatility and multiple periodicity of large-scale GPU workloads. Its core innovation is a dual-component encoder that combines primitive dictionary decomposition with adaptive spectral refinement, enabling precise capture of complex signal dynamics that monolithic models fail to represent.
    \item We introduce an interpretable decomposition paradigm that bridges the gap between black-box forecasting and actionable system insights. By disentangling aggregated workloads into learnable, semantically meaningful primitives, PRISM provides the granular transparency essential for risk-aware resource scheduling, surpassing the capabilities of traditional global fitting approaches.
    \item We conduct a comprehensive evaluation on a large-scale, real-world production trace from Alibaba Cloud. Experiments demonstrate that PRISM significantly outperforms state-of-the-art baselines, achieving a low MSE of 0.0753 and a high $R^2$-score of 0.9131.
\end{itemize}


\section{OVERVIEW OF PRISM}

\begin{figure*}[t]
\centering
\includegraphics[scale=0.43]{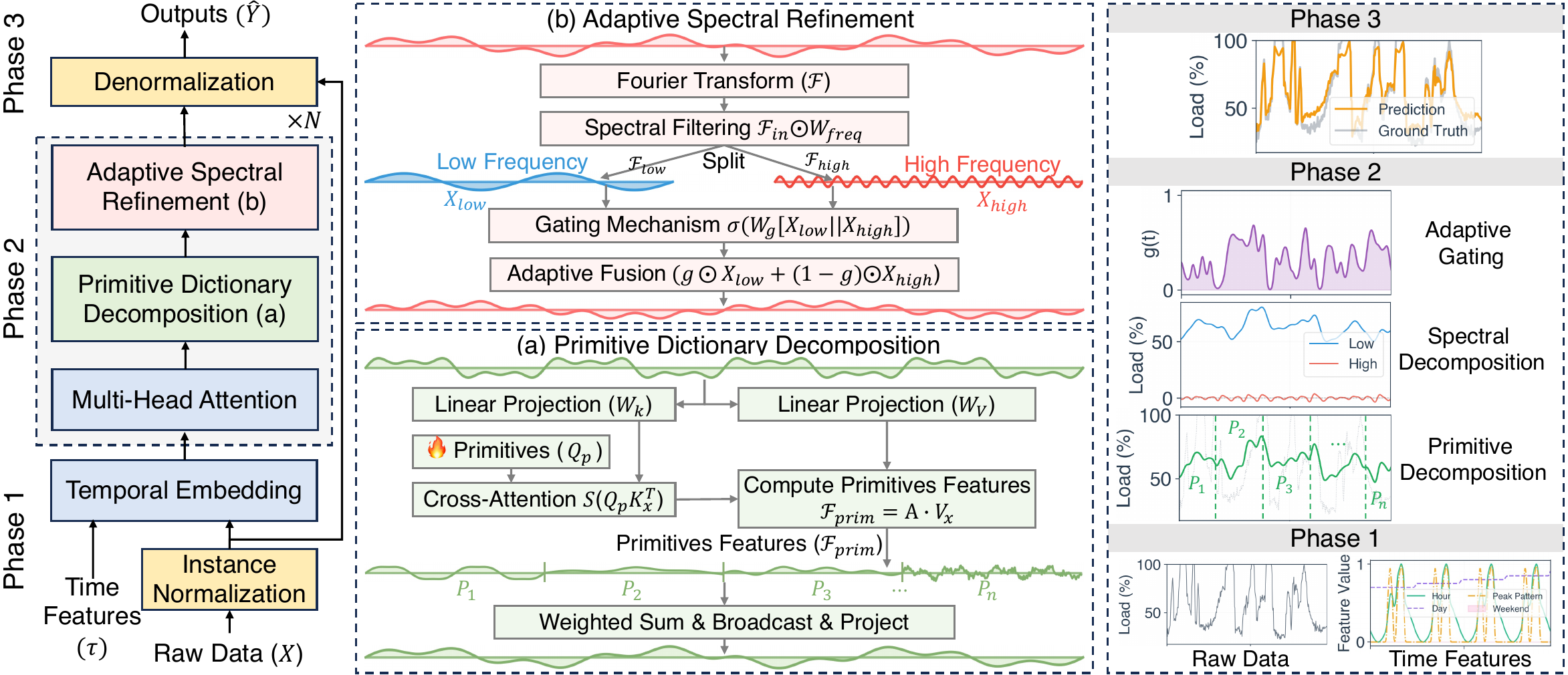} 
\caption{An overview of the PRISM framework.}
\label{Model}
\end{figure*}

As introduced in the prolegomenon, aggregated workloads in large-scale GPU clusters exhibit prominent non-stationarity and multi-component heterogeneity. This complexity arises from the superposition of numerous concurrent, heterogeneous micro-dynamics (e.g., job submissions with diverse SLAs, bursty data analytics tasks). Consequently, the conventional assumption of modeling the aggregated signal as a singular, homogeneous process is no longer tenable for automated resource orchestration and SLA assurance in modern data centers.

To address this challenge, we propose the PRISM framework. The central tenet of PRISM is to decompose the complex, seemingly stochastic aggregated signal into a set of simpler, interpretable, and dynamically combinable foundational primitive signals. This methodology is conceptually analogous to an optical prism, which decomposes white light into its constituent spectral components.

Fig. \ref{Model} presents an overview of the PRISM framework, whose execution flow is structured into three sequential phases:

\begin{enumerate}
 \item \textbf{Phase 1:} We construct an input representation that is robust to scale heterogeneity and rich in contextual information. The raw signal is first subject to instance-wise normalization to mitigate drastic statistical shifts. Subsequently, the normalized sequence is patched and deeply fused with multiple temporal features (e.g., intra-day, intra-week periodicities) to construct a high-dimensional temporal tensor.

\item \textbf{Phase 2:} We implement the core of our framework as an $N$-layer deep encoder. At each layer, we apply a synergistic dual-component process to the temporal tensor, achieving decomposability and dynamic modeling. This process consists of: (a) \textbf{Primitive Dictionary Decomposition}, where we dynamically decompose the sequence into a weighted combination of learnable, globally-shared primitive patterns, and (b) \textbf{Adaptive Spectral Refinement}, where we explicitly refine its representation in the frequency domain.

\item \textbf{Phase 3:} We feed the hierarchically refined and decomposed features into a prediction head. This head synthesizes the abstract representations and projects them onto the future forecast horizon $H$. We then optimize the model parameters via a composite objective function.
\end{enumerate}

In the following sections, we delve into the details of the PRISM framework.

\section{METHODOLOGY}

\subsection{Phase 1: Representation and Embedding}
\label{sec:Phase_1}

The workload forecasting task is formally defined as: given a historical observation sequence $X = \{x_1, \cdots, x_L\}$, where $x_t \in \mathbb{R}^{C}$ (with $C=1$ in this work), the goal is to learn a mapping $\mathcal{F}$ to predict the future $H$-step sequence $Y = \{x_{L+1}, \cdots, x_{L+H}\}$, such that $\hat{Y} = \mathcal{F}(X)$.

To handle the scale heterogeneity introduced by multi-tenancy and diverse job types, each input instance $X$ is independently normalized via instance-wise normalization:
\begin{align}
X_{\mathcal{N}} = \frac{X - \mu_X \mathbf{1}}{\sqrt{\sigma_X^2 + \epsilon}},
\end{align}
where $\mu_X = \frac{1}{L}\sum_{t=1}^L x_t$ and $\sigma_X^2 = \frac{1}{L}\sum_{t=1}^L (x_t - \mu_X)^2$ are the mean and variance of $X$ over the $L$ dimension, $\mathbf{1} \in \mathbb{R}^{L}$ is a vector of ones, and $\epsilon$ is a small constant for numerical stability.

Subsequently, $X_{\mathcal{N}}$ is partitioned into $N_p$ overlapping patches of length $P$ with stride $S$. Let $\Pop: \mathbb{R}^{L \times C} \to \mathbb{R}^{N_p \times (P \cdot C)}$ be the patching operator. We embed these patches into the $D$-dimensional space via a learnable linear projection $W_p \in \mathbb{R}^{(P \cdot C) \times D}$:
\begin{align}
X_p = \Pop(X_{\mathcal{N}}) W_p \in \mathbb{R}^{B \times N_p \times D},
\end{align}
where $B$ is the batch size. Concurrently, the corresponding timestamps $\{\tau_i\}_{i=1}^{N_p}$ are fed into a temporal embedding function $\mathcal{E}_t$ to generate the temporal representation $X_t = \mathcal{E}_t(\{\tau_i\}_{i=1}^{N_p}) \in \mathbb{R}^{B \times N_p \times D}$. The final output of Phase 1, $X^{(0)}$, is the element-wise sum of these two representations:
\begin{align}
X^{(0)} = X_p + X_t.
\end{align}

\subsection{Phase 2: Synergistic Decomposition and Spectral Refinement}
\label{sec:Phase_2}

This phase, which forms the core of our encoder, consists of a stack of $N$ identical layers. Each layer is designed to hierarchically decompose the input signal, refine its structure, and learn a robust representation of the input signal. Let $X^{(l-1)} \in \mathbb{R}^{B \times N_p \times D}$ be the input to the $l$-th layer, where $B$ is the batch size, $N_p$ is the sequence length, and $D$ is the model dimension. The processing flow, augmented with residual connections and layer normalization, proceeds as follows. First, a standard multi-head attention module captures the intra-sequence temporal dependencies:
\begin{align}
X'_{l} = \text{LN}\left(X^{(l-1)} + \text{MHA}(X^{(l-1)})\right).
\end{align}
$X'_{l}$ is then processed by two sequential sub-modules designed to model the compositional and spectral nature of time series explicitly.

\subsubsection*{\textbf{(a) Primitive Dictionary Decomposition}}

The objective of this module is to disentangle the complex input signal into a combination of simpler, semantically meaningful patterns. We introduce a learnable dictionary of $K$ primitives, $Q_p \in \mathbb{R}^{K \times D}$. Crucially, unlike standard latent embeddings, these primitives act as behavioral prototypes, effectively clustering recurrent workload dynamics (e.g., stable periodic cycles vs. transient bursty events) into distinct basis vectors. To achieve this, the input $X'_{l}$ is first projected into Key ($K_x = X'_{l}W_K$) and Value ($V_x = X'_{l}W_V$) matrices. A set of primitive features, $\mathcal{F}_{prim} \in \mathbb{R}^{B \times K \times D}$, is then extracted by computing scaled dot-product attention $A$ between the dictionary $Q_p$ (as Query) and the sequence's $K_x$ and $V_x$. This attention $A \in \mathbb{R}^{B \times N_H \times K \times N_p}$ is normalized over the sequence length $N_p$ (Eq. \ref{eq:pdd_a}), and the resulting $K$ primitive features are calculated via $A V_x$.
\begin{align}
\label{eq:pdd_a}
A = \text{Softmax}\left(\frac{Q_p K_x^T}{\sqrt{d_k}}\right).
\end{align}
Concurrently, a set of global selection weights, $\alpha \in \mathbb{R}^{B \times K}$, is computed to aggregate these primitives. This $\alpha$ is derived by averaging the pre-softmax attention logits across both heads ($N_H$) and sequence length ($N_p$), followed by a softmax over the $K$ primitives.
\begin{align}
\label{eq:pdd_alpha_s}
\bar{s}_{b,k} &= \frac{1}{N_H N_p} \sum_{h=1}^{N_H} \sum_{i=1}^{N_p} \left(\frac{Q_p K_x^T}{\sqrt{d_k}}\right)_{b, h, k, i}, 
\end{align}
\begin{align}
\alpha &= \text{Softmax}(\bar{s}).
\end{align}
These weights $\alpha$ serve as an explicit compositional recipe. By inspecting $\alpha$, we can transparently identify which primitive dominates the current timeframe, thereby explaining the prediction source (e.g., a spike driven by a burst primitive). This design intentionally decouples the global \textit{selection} of a primitive ($\alpha$) from the localized information extraction used to construct it ($A$). 

Crucially, to ensure this dictionary captures a rich and comprehensive set of patterns, we introduce a diversity-preservation objective inspired by OOD generalization principles. We posit that a robust dictionary must span the worst-case scenario of latent distributions. Therefore, we introduce an auxiliary loss, $\mathcal{L}_{div}$, during training. This loss maximizes the pairwise divergence (e.g., $\mathcal{H}$-divergence  or cosine distance) among the $K$ extracted primitive features $\{\mathcal{F}_{prim, k}\}_{k=1}^K$. Conceptually, this enforces semantic orthogonality, compelling each primitive to specialize in a unique, non-redundant sub-domain of the workload. This prevents mode collapse and ensures that the decomposed patterns remain distinct and human-readable. Finally, the aggregated feature vector $h_{agg} = \sum_{k=1}^K \alpha_k \cdot \mathcal{F}_{prim, k}$ is broadcast to match the sequence length and injected back into the sequence via a residual connection:
\begin{align}
X''_{l} = \text{LN}\left(X'_{l} + \text{Broadcast}(h_{agg}) W_O\right).
\end{align}

\begin{algorithm}
\DontPrintSemicolon
\caption{Training Procedure of PRISM}
\label{alg:prism}
\KwInput{Training data $\mathcal{D}=\{(X_i, Y_i)\}_{i=1}^{M}$; model parameters $\Theta$; hyperparameters $N, K, \lambda_1, \lambda_{div}$.}
\KwOutput{Trained model parameters $\Theta^*$.}

\For{each training iteration until convergence}{
    Randomly sample mini-batch $\{X, Y\}$ from $\mathcal{D}$\;
    $X_{\mathcal{N}}, \mu_X, \sigma_X \leftarrow \text{NormalizeInstance}(X)$\;
    $X_p \leftarrow \text{PatchEmbed}(X_{\mathcal{N}})$ \tcp{Patch and embed}
    $X_t \leftarrow \mathcal{E}_t(\{\tau_i\})$ \tcp{Time embedding}
    Initialize encoder input $X^{(0)} \leftarrow X_p + X_t$\;
    $\mathcal{L}_{div\_total} \leftarrow 0$\;
    
    \For{$l = 1$ \KwTo $N$}{
        $X'_{l} \leftarrow \text{LN}(X^{(l-1)} + \text{MHA}(X^{(l-1)}))$\;
        \tcp{Primitive Dictionary Decomposition}
        $K_x \leftarrow X'_l W_K$, $V_x \leftarrow X'_l W_V$\;
        $A \leftarrow \text{Softmax}(Q_p K_x^T / \sqrt{d_k})$ \tcp{Local attention}
        $\mathcal{F}_{\text{prim}} \leftarrow \text{Concat}_h(A_h V_h)$ \tcp{K primitives}
        $\alpha \leftarrow \text{Softmax}(\text{Mean}_{h,i}(Q_p K_x^T / \sqrt{d_k}))$ \tcp{Global}
        $h_{agg} \leftarrow \sum_{k=1}^{K} \alpha_k \cdot \mathcal{F}_{\text{prim},k}$\;
        $X''_{l} \leftarrow \text{LN}(X'_{l} + \text{Broadcast}(h_{agg}) W_O)$\;
        $\mathcal{L}_{div}^{(l)} \leftarrow \text{ComputeDiversityLoss}(\mathcal{F}_{\text{prim}})$\;
        $\mathcal{L}_{div\_total} \leftarrow \mathcal{L}_{div\_total} + \mathcal{L}_{div}^{(l)}$\;
        
        \tcp{Adaptive Spectral Refinement}
        $\mathcal{F}_{in} \leftarrow \mathcal{F}(X''_{l})$\;
        $\mathcal{F}_w \leftarrow \mathcal{F}_{in} \odot W_{\text{freq}}$ \tcp{Complex Weight Filtering}
        $\mathcal{F}_{low}, \mathcal{F}_{high} \leftarrow \text{MaskSplit}(\mathcal{F}_w)$ \tcp{Freq Masking}
        $X_{low} \leftarrow \mathcal{F}^{-1}(\mathcal{F}_{low}) W_{low}$\; 
        $X_{high} \leftarrow \mathcal{F}^{-1}(\mathcal{F}_{high}) W_{high}$\;
        $g \leftarrow \sigma(W_g[X_{low} \mathbin\Vert X_{high}] + b_g)$ \tcp{Gating}
        $X'''_l \leftarrow \text{LN}(X''_l + (g \odot X_{low} + (1-g)\odot X_{high}))$\;
        $X^{(l)} \leftarrow \text{LN}(X'''_l + \text{FFN}(X'''_l))$\;
    }
    
    $h_c \leftarrow \text{MeanPooling}(X^{(N)})$, $\hat{Y}_{\mathcal{N}} \leftarrow \text{MLP}(h_c)$\;
    $\hat{Y} \leftarrow \text{Denormalize}(\hat{Y}_{\mathcal{N}}, \mu_X, \sigma_X)$\;
    $\mathcal{L}_{pre} \leftarrow \mathbb{E}[\|\hat{Y}-Y\|_2^2] + \lambda_1 \cdot \mathbb{E}[\|\hat{Y}-Y\|_1]$\;
    $\mathcal{L}_{total} \leftarrow \mathcal{L}_{pre} + \lambda_{div} \cdot \mathcal{L}_{div\_total}$\;
    $\Theta \leftarrow \text{OptimizerStep}(\Theta, \nabla_\Theta \mathcal{L}_{total})$\;
}
\KwRet{$\Theta^*$}
\end{algorithm}

\subsubsection*{\textbf{(b) Adaptive Spectral Refinement}}

The motivation for this module is to explicitly refine the signal's representation in the frequency domain, enabling the model to learn its complex spectral properties adaptively. The signal $X''_{l}$ undergoes a filter-and-refine process. It is first projected into the frequency domain via a real fast Fourier Transform ($\mathcal{F}$), yielding $\mathcal{F}_{in} = \mathcal{F}(X''_{l}) \in \mathbb{C}^{B \times F \times D}$, where $F$ is the number of frequency components. An adaptive filtering step is then performed, where a learnable frequency-wise weight filter $W_{freq} \in \mathbb{C}^{F \times D}$ is applied via the element-wise product: 
\begin{equation}
\mathcal{F}_{w} = \mathcal{F}_{in} \odot W_{freq}.
\end{equation}
The filtered signal $\mathcal{F}_{w}$ is decomposed into low-frequency $\mathcal{F}_{low}$ and high-frequency $\mathcal{F}_{high}$ components via a frequency masking strategy (zeroing out components beyond/below a cutoff) to preserve the temporal scale. These components are transformed back to the time domain via the inverse FFT ($\mathcal{F}^{-1}$) and passed through independent linear layers $W_{low}$ and $W_{high}$ to obtain $X_{low}$ and $X_{high}$, respectively. A dynamic gating mechanism is then employed to fuse these two refined components. A gate $g$ is computed based on the concatenation of both components (Eq. \ref{eq:asr_g}), which in turn determines their weighted contribution to the final output (Eq. \ref{eq:asr_out}).
\begin{align}
\label{eq:asr_g}
g &= \sigma\left(W_g[X_{low} \mathbin\Vert X_{high}] + b_g\right),
\end{align}
\begin{align}
\label{eq:asr_out}
X'''_{l} &= \text{LN}\left(X''_{l} + (g \odot X_{low} + (1-g) \odot X_{high})\right).
\end{align}
Finally, $X'''_{l}$ passes through a standard position-wise feed-forward network (FFN) to produce the layer's final output $X^{(l)}$:
\begin{align}
X^{(l)} = \text{LN}\left(X'''_{l} + \text{FFN}(X'''_{l})\right).
\end{align}

\subsection{Phase 3: Predictive and Optimization}
\label{sec:Phase_3}

After $N$ refinement layers, the final output $X^{(N)} \in \mathbb{R}^{B \times N_p \times D}$ encapsulates the high-dimensional representation of the sequence. For forecasting, we first extract a consolidated context vector $h_c \in \mathbb{R}^{B \times D}$ by applying a permutation-invariant operator (mean pooling) over the time dimension ($N_p$):
\begin{align}
h_c = \text{MeanPooling}(X^{(N)}) = \frac{1}{N_p} \sum_{i=1}^{N_p} X^{(N)}_{[:, i, :]}.
\end{align}
This fixed-size representation is then fed into a multi-layer perceptron head to project it to the $H$-step forecast $\hat{Y}_{\mathcal{N}} \in \mathbb{R}^{B \times H}$: $\hat{Y}_{\mathcal{N}} = \text{MLP}(h_c)$.
$\hat{Y}_{\mathcal{N}}$ is subsequently denormalized (per Phase 1) to produce the final prediction $\hat{Y}$.

The model parameters $\Theta$ are optimized by minimizing a joint objective function $\mathcal{L}_{total}(\Theta)$. This objective integrates the primary forecasting error with the auxiliary diversity regularization from Phase 2 to ensure both predictive accuracy and representation robustness:
\begin{align}
\mathcal{L}_{pre} &= \mathbb{E}_{X,Y}\left[\|\hat{Y} - Y\|_2^2\right] + \lambda_1 \cdot \mathbb{E}_{X,Y}\left[\|\hat{Y} - Y\|_1\right], 
\end{align}
\begin{align}
\mathcal{L}_{total}(\Theta) &= \mathcal{L}_{pre} + \lambda_{div} \cdot \sum_{l=1}^{N} \mathcal{L}_{div}^{(l)},
\end{align}
where $\mathcal{L}_{forecast}$ combines mean squared error (MSE) and mean absolute error (MAE), and $\mathcal{L}_{div}^{(l)}$ is the diversity-preservation loss computed in the $l$-th encoder layer (as defined in Phase 2). $\lambda_1$ and $\lambda_{div}$ are hyperparameters balancing the loss components.

\subsection{Complexity and Latency Analysis}
We analyze the efficiency using $N$ layers, $N_p$ patches where $N_p \ll L$, dimension $D$, and $K$ primitives. The \textbf{time complexity} scales as $O(N(N_p^2 D + N_p D^2))$, effectively reducing the quadratic dependency from $O(L^2)$ to $O(N_p^2)$ to achieve the millisecond-level inference latency essential for real-time scheduling. In terms of \textbf{space complexity}, the training phase requires $O(B N_H N_p^2)$ memory for attention maps, whereas the inference phase is bounded only by the model parameters $O(N(D^2 + KD))$. This efficiency facilitates lightweight deployment on cluster management nodes.

The overall procedure of the proposed PRISM algorithm is outlined in Algorithm \ref{alg:prism}.

\section{EXPERIMENTAL RESULTS}

To demonstrate the practical effectiveness and interpretability of PRISM for forecasting GPU cluster workloads, we utilize a large-scale, real-world trace dataset from an Alibaba production GPU cluster. This dataset chronicles the complete lifecycle of 466,867 AI tasks, spanning 184.07 days from April to October 2024. The cluster is equipped with 10,412 GPUs (e.g., A10, A100, and A800); detailed resource configuration and allocation are presented in Table \ref{tab_GPUs}.

We present PRISM against 9 SOTA baselines: Autoformer, Dlinear, Fedformer \cite{zhou2022fedformer}, Informer \cite{2021AAAIInformer}, TimesNet, TimeXer \cite{wang2024timexer}, WpMixer \cite{murad2025wpmixer}, MetaEformer, and Orglinear \cite{duan2025gfs}. We compared performance across multiple prediction horizons using standard metrics, including MSE, MAE, Root Mean Squared Error (RMSE), and $R^2$.


\begin{table}
\centering
\caption{GPU resource configuration and allocation overview.}
\label{tab_GPUs}
\renewcommand{\arraystretch}{1.2}
\setlength{\tabcolsep}{5pt}
\begin{tabular}{@{}l c c c c}
\toprule
GPU Model & Config. & Nodes & GPUs &Alloc. Rate \\ 
\midrule
A10 & 1×GPU & 2,494 & 2,494 & \redcircle \\
A100-SXM4-80GB & 8×GPU & 432 & 3,456 & \greencircle \\
A800-SXM4-80GB & 8×GPU & 22 & 176 & \greencircle \\
GPU-series-1 & Mixed & 989 & 1,558 & \yellowcircle \\
GPU-series-2 & 8×GPU & 122 & 976 & \bluecircle \\
H800 & 8×GPU & 219 & 1,752 & \bluecircle \\
\bottomrule
\end{tabular}
\vspace{0.5em}
\noindent
\parbox{\linewidth}{%
\setlength{\parindent}{0pt} 
\setlength{\parskip}{0pt}   
Note: GPU-series-1 is a mixed cluster (902 x 1-GPU nodes, 77 x 8-GPU nodes, 10 x 4-GPU nodes). Utilization levels: \redcircle\ Over-subscribed ($>$100\%); \yellowcircle\ Near-capacity (70-99\%); \greencircle\ Moderate (40-69\%); \bluecircle\ Underutilized ($<$40\%).
}
\end{table}

%
%

\subsection{Prediction Results}

\subsubsection*{\textbf{Overall Performance Comparison}}

\begin{figure*}[t]
\centering
\includegraphics[width=0.9\textwidth]{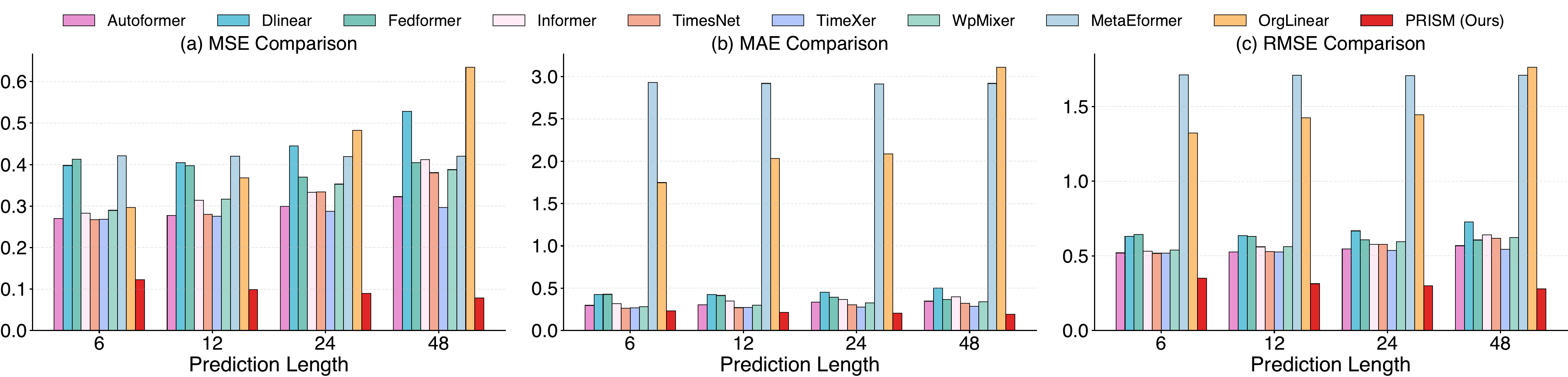}
\caption{Main forecasting results of PRISM and baseline models at prediction lengths of 6, 12, 24, and 48.}
\label{fig:results_main_all}
\end{figure*}

\begin{figure}
    \centering
    \includegraphics[width=\columnwidth]{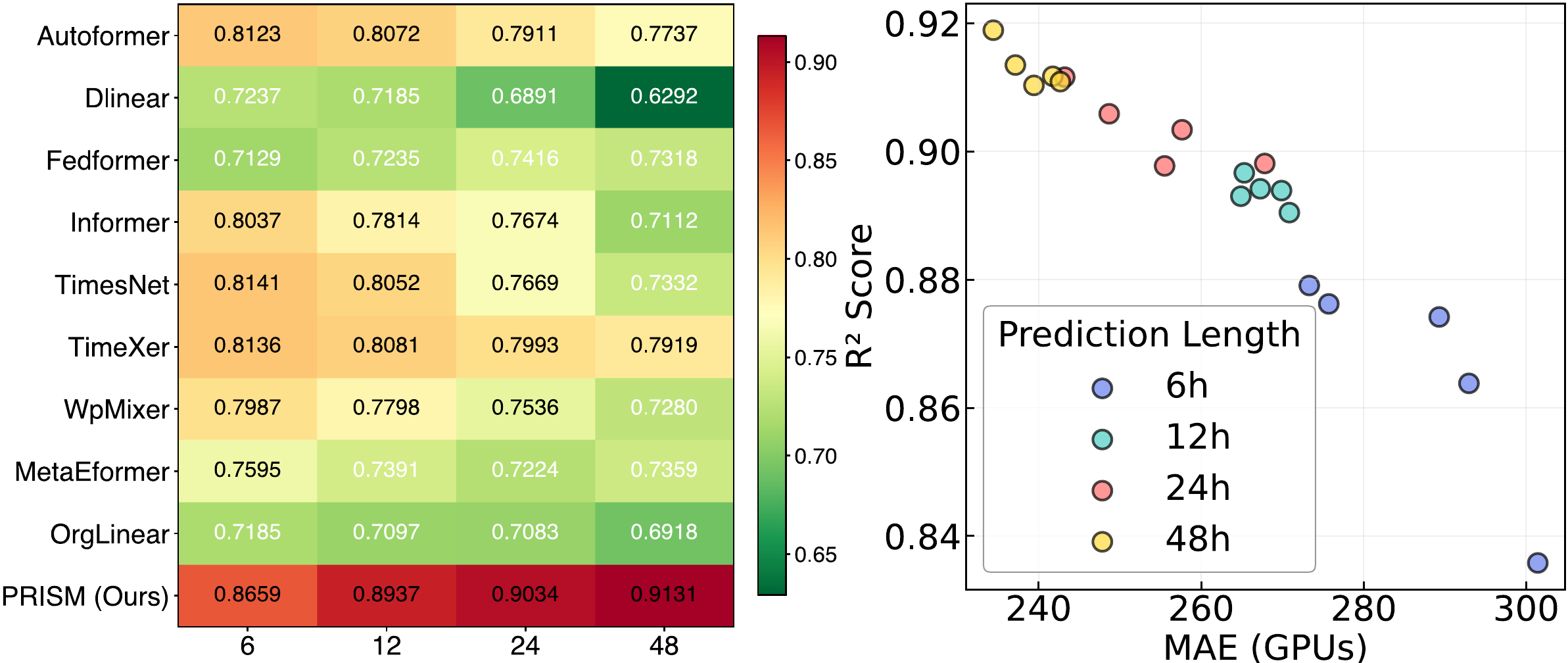}
        \caption{Performance comparison of PRISM against baselines.}
    \label{fig:r2_all} 
\end{figure}

We conducted a comprehensive quantitative comparison of PRISM against nine state-of-the-art baselines. As summarized in Fig. \ref{fig:results_main_all} and Fig. \ref{fig:r2_all}, PRISM consistently and significantly outperforms all competing methods across all evaluation metrics and prediction horizons. This overarching superiority is detailed in the following breakdown of error magnitude and goodness-of-fit.

First, in terms of error magnitude and long-term robustness (Fig. \ref{fig:results_main_all}), PRISM achieves dramatically lower errors than all baselines. In Fig. \ref{fig:results_main_all}a, where large deviations are heavily penalized, PRISM consistently avoids catastrophic failures, unlike models such as OrgLinear whose MSE values are an order of magnitude higher. This behavior ensures stable system performance even when handling the extreme peaks shown in Fig. \ref{fig_intro_2}. A similar pattern appears in Fig. \ref{fig:results_main_all}b, where the gap between PRISM and the baselines grows as the prediction length increases. At the 48-hour horizon, PRISM maintains exceptionally low and stable errors that are barely visible compared with the high bars of OrgLinear and TimeXer, which highlights its robustness in scenarios where other baselines lose reliability.

Second, regarding goodness-of-fit (Fig. \ref{fig:r2_all}), PRISM further solidifies its dominance from an $R^2$ perspective. The heatmap (left) confirms that PRISM achieves the highest $R^2$ score across all horizons, with a unique trend of improvement at longer horizons, peaking at 0.9131 at 48h. The scatter plot (right) contextualizes this performance, showing that PRISM occupies the ideal top-left quadrant (lowest MAE, highest $R^2$) for all prediction lengths, distinguishing it clearly from the competing methods.

\subsubsection*{\textbf{Ablation Study}}
As shown in Table \ref{tab:prism_ablation_percent}, our ablation study confirms the necessity of PRISM's components. Focusing on the primitive and spectral, we observe that removing either one (w/o-primitive or w/o-spectral) degrades performance. Critically, removing both (w/o-prim-spec) results in a disproportionately severe performance drop of 15.27\% (MSE). This degradation, which far exceeds the sum of their individual impacts, suggests a powerful synergistic effect between the primitive and spectral, indicating that their combination is crucial for the model's high accuracy.

%
\begin{figure}[t]
    \centering 
    \begin{subfigure}{\columnwidth}
        \centering
        \includegraphics[width=0.85\linewidth]{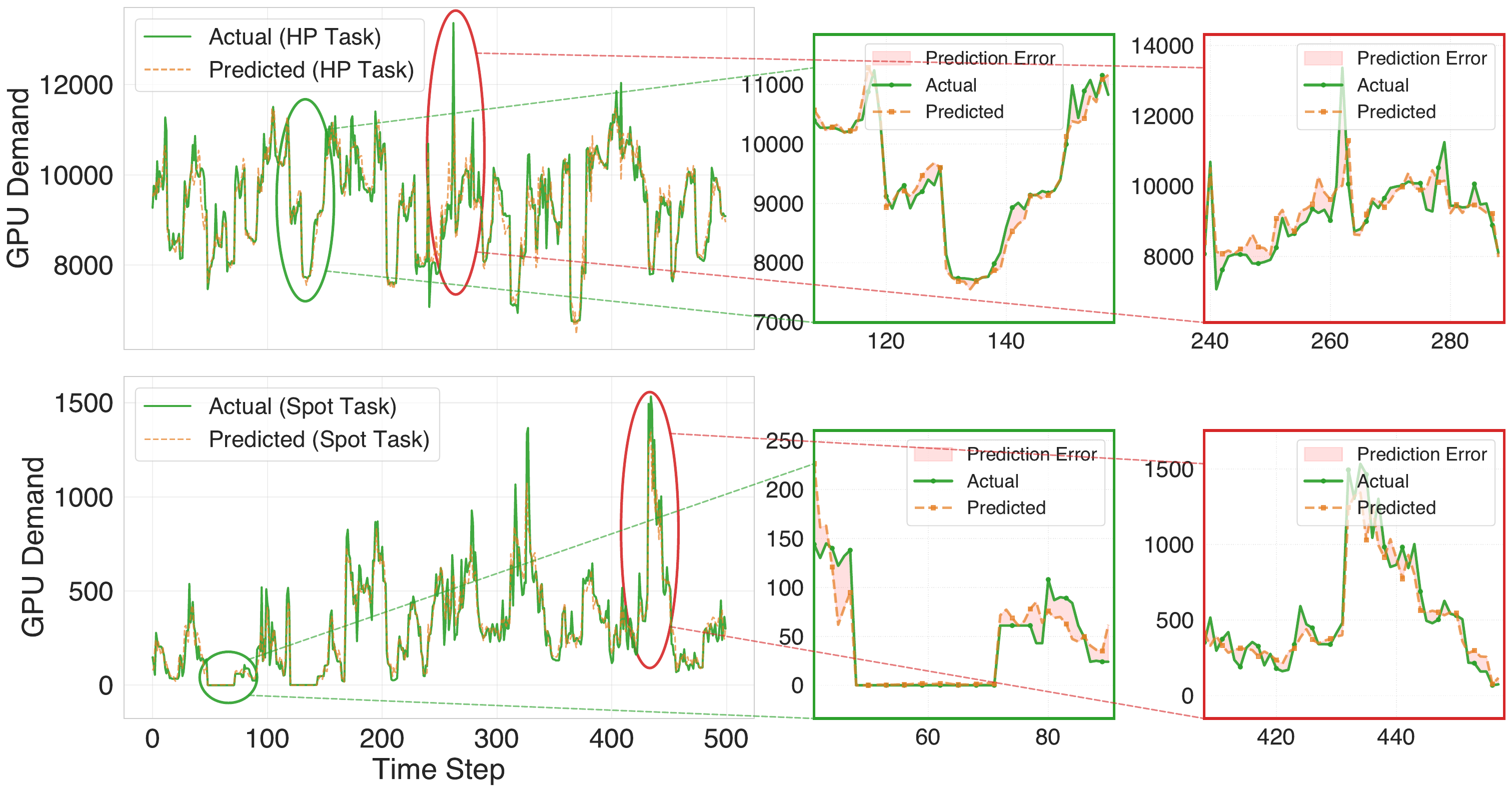}
        \caption{Fine-grained prediction for HP and Spot tasks.} 
        \label{fig:hp_vs_spot_predictions_zoom}
    \end{subfigure}
    \begin{subfigure}{\columnwidth}
        \centering
        \includegraphics[width=0.85\linewidth]{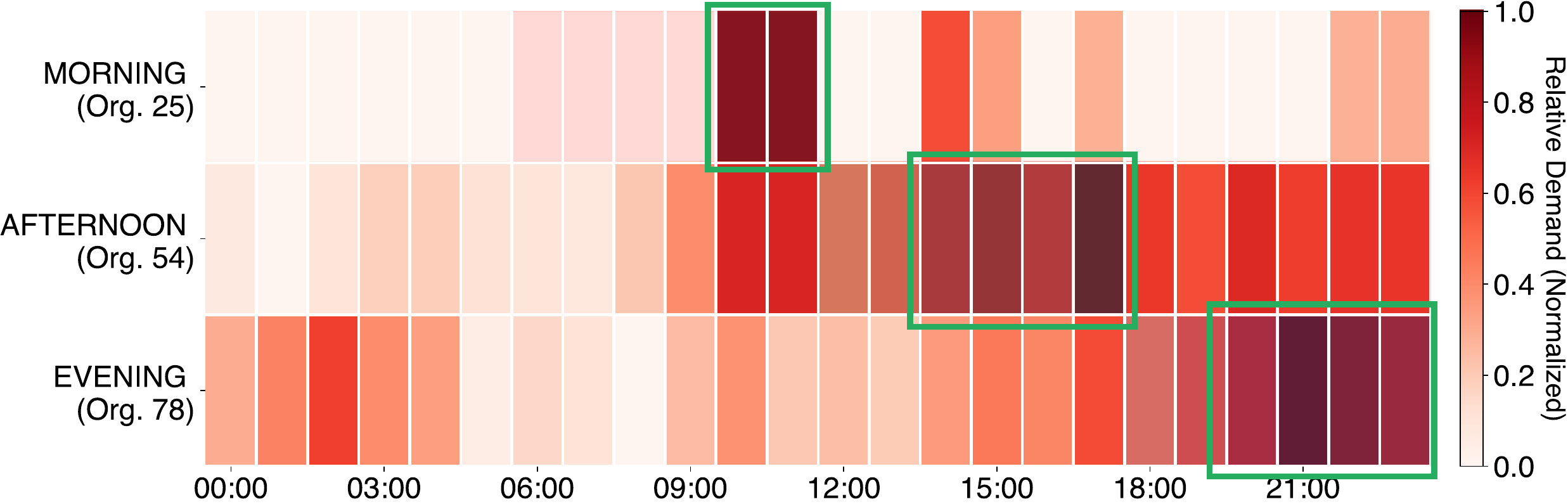}
        \caption{Diurnal patterns of three organizations.} 
        \label{fig:24hour_prediction_heatmap_3_ORGS}
    \end{subfigure}
    \caption{Case study on large-scale GPU cluster workloads.}
    \label{fig:combined_results}
\end{figure}



\begin{table}
\centering
\setlength{\tabcolsep}{1.2pt} 
\caption{PRISM ablation model performance, as a percentage change relative to the full model (first row).}
\label{tab:prism_ablation_percent}
\begin{tabular}{l ccc cccc}
\toprule
Methods & Pat. & Prim. & Spec. & MSE ($\downarrow$) &MAE ($\downarrow$) & RMSE ($\downarrow$) &R$^2$($\uparrow$) \\
\midrule
PRISM &  \Checkmark &  \Checkmark &  \Checkmark & 0.0753 & 0.1926 & 0.2744 & 0.9131 \\
w/o-patch & \textcolor[cmyk]{0,0,0,.247}  {\XSolidBrush} &  \Checkmark &  \Checkmark & +11.82\%&+5.87\% &+5.72\% & -1.09\% \\
w/o-primitive &  \Checkmark & \textcolor[cmyk]{0,0,0,.247}  {\XSolidBrush} &  \Checkmark & +1.33\% & +0.52\% & +0.62\% &-0.12\% \\
w/o-spectral &  \Checkmark &  \Checkmark &\textcolor[cmyk]{0,0,0,.247}  {\XSolidBrush} &+3.19\% & +1.77\% & +1.57\% &-0.29\% \\
w/o-prim-spec & \Checkmark& \textcolor[cmyk]{0,0,0,.247}  {\XSolidBrush} &\textcolor[cmyk]{0,0,0,.247}  {\XSolidBrush} &+15.27\% &+5.66\% & +0.95\% &-0.19\% \\
Baseline &\textcolor[cmyk]{0,0,0,.247}  {\XSolidBrush} &\textcolor[cmyk]{0,0,0,.247}  {\XSolidBrush} &\textcolor[cmyk]{0,0,0,.247}  {\XSolidBrush}& +21.91\% &+10.64\% &+10.42\% &-2.03\% \\
\bottomrule
\end{tabular}
\end{table}


\subsection{Case Study}
To qualitatively validate PRISM's superior performance, our case study in Fig. \ref{fig:combined_results} visually corroborates the data challenges identified in Fig. \ref{fig_intro_2}. Fig. \ref{fig:hp_vs_spot_predictions_zoom} demonstrates that PRISM successfully captures task-level heterogeneity (from Fig. \ref{fig_intro_2}c), achieving high-fidelity predictions by accurately tracking sharp peaks for both volatile HP tasks and bursty Spot tasks. Concurrently, Fig. \ref{fig:24hour_prediction_heatmap_3_ORGS} reveals its ability to model the diverse organizational patterns (from Fig. \ref{fig_intro_2}b and Fig. \ref{fig_intro_2}c), such as the distinct "morning peak" vs. "afternoon peak", successfully avoiding the pitfall of fitting a non-existent "average user". This simultaneous ability to precisely model heterogeneous tasks and organizations is key to PRISM's robust forecasting accuracy.

\section{CONCLUSION}
In this work, we introduce PRISM, a novel compositional framework for managing the volatility, periodicity, and heterogeneity of large-scale GPU workloads. By decomposing the signal into a dynamic mixture of interpretable primitives, PRISM overcomes the limitations of monolithic models. Comprehensive experiments on a real-world dataset validated PRISM's state-of-the-art performance and its ability to model heterogeneous patterns via its synergistic core components. PRISM provides a robust and interpretable tool for automated resource management in modern AI infrastructure.

\section*{ACKNOWLEDGMENTS}
This work was supported by the National Natural Science Foundation of China (No. 62272398) and the Sichuan Science and Technology Program (No. 2024NSFJQ0019).



\bibliographystyle{ACM-Reference-Format}
\bibliography{references}

%
%
%
%
%
%
%
%

\end{document}